\documentclass[doublecol,figures]{epl2}

\usepackage{epsfig}
\usepackage{amsmath}
\usepackage{amsfonts}
\usepackage{amssymb}
\usepackage{graphicx}
\usepackage{units}
\usepackage{url}
\usepackage{color}

\newcommand{\be}{\begin{equation}}
\newcommand{\ee}{\end{equation}}
\newcommand{\bmul}{\begin{multline}}
\newcommand{\emul}{\end{multline}}
\newcommand{\bea}{\begin{eqnarray}}
\newcommand{\eea}{\end{eqnarray}}
\newcommand{\rr}{\mathbf{r}}
\newcommand{\kk}{\mathbf{k}}
\newcommand{\qq}{\mathbf{q}}
\newcommand{\vv}{\mathbf{v}}

\newcommand{\eeee}{\mathbf{e}}

\newcommand{\bra}[1]{\langle #1|}
\newcommand{\ket}[1]{|#1\rangle}

\newcommand{\bb}[1]{\left( #1 \right)}

\newcommand{\bbcro}[1]{\left[ #1 \right]}

\newcommand{\ii}{\textrm{i}}
\newcommand{\eee}{\textrm{e}}
\newcommand{\dd}{\mathrm{d}}

\DeclareMathOperator\argsh{argsh}

\DeclareMathOperator\re{Re}

\title{Landau-Khalatnikov phonon damping in strongly interacting Fermi gases}
\shorttitle{Landau-Khalatnikov phonon damping in strongly interacting Fermi gases}
\author{Hadrien Kurkjian, Yvan Castin, Alice Sinatra}
\shortauthor{H. Kurkjian, Y. Castin, A. Sinatra}
\institute{Laboratoire Kastler Brossel, ENS-PSL, CNRS, UPMC-Sorbonne Universit\'es and Coll\`ege de France, Paris, France}


\abstract{We derive the phonon damping rate due to the four-phonon  Landau-Khalatnikov process in low temperature strongly interacting Fermi gases using quantum hydrodynamics, correcting and extending the original calculation of Landau and Khalatnikov [ZhETF, {\bf 19} (1949) 637]. Our predictions can be tested in state-of-the-art experiments with cold atomic gases in the collisionless regime.}

\begin{document}
\date{6 October 2016}
\maketitle

\section{Introduction}

Phonons, sound waves, low energy normal modes or gapless collective excitations are ubiquitous in physics. In uniform 
weakly-excited quantum many-body systems with short-range interactions, they are described as quasiparticles characterized by a dispersion relation approximately linear at low wavenumber, $\omega_\qq{\sim}cq$ with $c$ the speed of sound, and by a damping rate much smaller than the angular eigenfrequency $\Gamma_\qq\ll \omega_\qq$. 
Phonon damping plays a central role in transport phenomena such as thermal conduction in dielectric solids, and in hydrodynamic properties such as temperature dependent viscosity and attenuation of sound in liquid helium \cite{Khalatnikov1949,Khalatnikov1966}. It is also crucial for macroscopic coherence properties, since it determines the intrinsic coherence time of bosonic and fermionic gases in the condensed or pair-condensed regime \cite{Zoller1998,CastinSinatra2009,CRP2016}.   
In the absence of impurities the damping of low-energy phonons is determined by phonon-phonon interactions that conserve energy and momentum and it crucially depends on the curvature of the phonon dispersion relation \cite{Wyatt1992,Wyatt2009}. For a concave dispersion relation, $1\leftrightarrow 2$ 
Beliaev-Landau processes involving three phonons are not resonant and the $2 \leftrightarrow 2$ Landau-Khalatnikov process involving four quasiparticles dominates at low $q$. 

In this paper we consider an unpolarized gas of spin-$1/2$ fermions prepared in thermal equilibrium at a temperature $T$ below the critical temperature, where a macroscopic coherence between pairs of opposite spin fermions appears.
Compared to other many-body fermionic systems, atomic gases offer the unique possibility to tune the interaction strength with an external magnetic field close to a so-called Feshbach resonance. This allows experimentalists to
explore the crossover between the Bose-Einstein Condensate (BEC) and Bardeen-Cooper-Schrieffer (BCS) 
regimes \cite{Thomas2002,Salomon2003,Grimm2004,Grimm2004b,Ketterle2004,Ketterle2005,Salomon2010,Zwierlein2012,Grimm2013}. The dispersion relation of low energy excitations, describing the collective motion of the pair center of mass, has a phononic start at small wavenumbers \cite{Anderson1958,Strinati1998,CKS2006,Tempere2011,Randeria2014,KCS2016} and changes from convex to concave in the BEC-BCS crossover, close to the strongly interacting unitary limit \cite{Tempere2011,KCS2016}. Therefore, the damping caused by the $2 \leftrightarrow 2$  processes should be directly observable in cold Fermi gases, contrarily to weakly-interacting Bose gases where the convex Bogoliubov dispersion relation supports Landau-Beliaev damping. On the theoretical side, the original study by Landau and Khalatnikov of the $2\leftrightarrow2$ damping rate \cite{Khalatnikov1949} is limited to the case where one of the colliding phonons has a small wavenumber compared to the other and it performs as we shall see an unjustified approximation on the coupling amplitude. 
Here, we give the general expression of the phonon damping rate in the concave dispersion relation regime at low temperature, where it is dominated by the $2 \leftrightarrow 2$ processes, correcting and extending the original calculation of reference \cite{Khalatnikov1949}.
In the whole paper we restrict to the so-called collisionless regime
where the phonon angular frequency times the typical collision time in the gas is much larger than one, $\omega_\qq \tau_c \gg 1$
\cite{Ketterle1998a,Walraven2005}. 
This is in general the case in superfluid gases at low temperature 
\footnote{One can estimate $\tau_c \simeq 1/\Gamma_{\qq_{\rm th}}$ where 
$\hbar c q_{\rm th} = k_BT$. Then for excitation frequencies scaling 
as $k_BT$, as in eq.~\eqref{eq:eps_tildeq}, the condition $\omega_\qq \gg \Gamma_\qq$, satisfied for a weakly-excited gas, implies $\omega_\qq \gg \Gamma_{\qq_{\rm th}}$ which ensures the collisionless regime.}. 

\section{Effective $2\leftrightarrow 2$ phonon coupling}

The theoretical framework we use is the irrotational quantum hydrodynamics of Landau and Khalatnikov \cite{Khalatnikov1949}.
Quantum hydrodynamics is an effective low-energy theory that relies only on the equation of state and can thus be
applied in all interaction regimes. For fermions, it neglects from the start the internal fermionic degrees of freedom
and the corresponding gapped BCS excitation spectrum,
treating the pairs of fermions at large spatial scales as a bosonic field. It is expected to give exact results to leading
order in the low temperature $T$, at least for observables involving low energy scales.
The quantum hydrodynamics Hamiltonian\footnote{In principle, this Hamiltonian has to be regularized by introducing an ultraviolet momentum cut-off or by discretizing the real space on a lattice as in reference \cite{CRP2016}. This however does not play a role here.} reads
\begin{equation}
\hat{H} = \int \dd^3r \left[\frac{\hbar^2}{2m} \nabla \hat{\phi}\cdot \hat{\rho}\ \nabla \hat{\phi}
+ e_{0}(\hat{\rho})\right]
\label{eq:hamiltonienhydro}
\end{equation}
where $e_{0}(\hat{\rho})$ is the ground state energy density, $m$ is the mass of a particle, and the superfluid velocity 
field operator $\hat{\vv}(\rr,t)=\frac{\hbar}{m}\nabla \hat{\phi}(\rr,t)$ is the gradient of the phase field operator, canonically conjugated to the density field operator $\hat{\rho}(\rr,t)$, so that $[\hat{\rho}(\rr,t),\hat{\phi}(\rr',t)]=\ii \delta(\rr-\rr')$. 
Assuming small deviations of 
$\hat{\rho}(\rr,t)$ and $\hat{\phi}(\rr,t)$ from their uniform spatial averages, one finds
the normal modes Fourier components $\delta{\hat{\rho}}_\qq \propto q^{1/2}(\hat{b}_\qq+\hat{b}_{-\qq}^\dagger)$ and
$\delta{\hat{\phi}}_\qq \propto -\ii q^{-1/2}(\hat{b}_\qq-\hat{b}_{-\qq}^\dagger)$, where the annihilation and 
creation operators 
of quasiparticles $\hat{b}_\qq$ and  $\hat{b}_\qq^\dagger$ obey bosonic commutation relations.
One inserts the expansion of $\hat{\rho}$ and $\hat{\phi}$ over these
modes in the Hamiltonian \eqref{eq:hamiltonienhydro}, which results in the series $\hat{H}=E_0+\hat{H}_2+\hat{H}_3+\hat{H}_4+\ldots$\,, where the index refers to the total degree in $\hat{b}_\qq$ and $\hat{b}_\qq^\dagger$,
each term being written in normal order.

In the concave dispersion relation regime considered in this paper, 
the direct phonon coupling due to $\hat{H}_3$ is not resonant, and the leading resonant coupling is a four-phonon
process. An effective interaction Hamiltonian $\hat{H}_{\rm eff}$ coupling an initial Fock state of quasiparticles $|i\rangle$ of energy
$E_i$ to a final one $|f\rangle$ of same energy $E_f=E_i$, 
where two wavevectors $\qq_1$ and $\qq_2$ are annihilated and 
two other wavevectors $\qq_3$ and $\qq_4$ are created, can then be derived in second order perturbation theory by considering the direct coupling by 
$\hat{H}_4$ to first order and the indirect coupling (involving a non-resonant intermediate state $|\lambda\rangle$) by $\hat{H}_3$ to second order, 
\be
\bra{f} \hat{H}_{\rm eff} \ket{i} \simeq  \bra{f} \hat{H}_4 \ket{i}+ \sum_\lambda \frac{\bra{f} \hat{H}_3 \ket{\lambda} \bra{\lambda} \hat{H}_3 \ket{i}}{E_i-E_\lambda}\equiv\mathcal{A}_{i\to f}
\label{eq:Aif}
\ee
The reader will notice that, for the purely linear dispersion relation $\omega_\qq =cq$ predicted by $\hat{H}_2$, the denominators in \eqref{eq:Aif} vanish for aligned wavevectors because the intermediate processes become resonant, for example $E_i-E_\lambda=\omega_1+\omega_2 -\omega_{\qq_1+\qq_2}=0$,
 with the short-hand notation $\omega_i \equiv \omega_{\qq_i}$. Following Landau and Khalatnikov, we regularize 
the resulting divergence in $\mathcal{A}_{i\to f}$ by including the actual curvature of the spectrum  \cite{Tempere2011,KCS2016} in the energy denominators\footnote{The quantum
 hydrodynamics Hamiltonian can be supplemented by terms leading to a curved dispersion relation \cite{Salasnich2015}. Except 
 in the energy denominator, this brings a negligible correction to the phonon damping rate at low temperature.}
\be
\hbar \omega_\qq \underset{q\to 0}{=} \hbar c q \left[1+\frac{\gamma}{8} \bb{\frac{\hbar q}{mc}}^2+ 
O\bb{\frac{\hbar q}{mc}}^4
\right]. \label{eq:fleur}
\ee
Here the speed of sound $c$ is related to the gas density $\rho$ and the ground state chemical potential $\mu$ by
\be
m c^2 = \rho \frac{\dd \mu}{\dd\rho}
\ee
whereas the dimensionless curvature parameter $\gamma<0$ must be measured or determined from a microscopic theory.
By introducing the dimensionless and state-independent effective coupling amplitude $\mathcal{A}_{\rm eff}$,
\be
\mathcal{A}_{i\to f} = \sqrt{n_{\qq_1} n_{\qq_2} (1+n_{\qq_3})(1+n_{\qq_4})} \frac{4mc^2}{\rho L^3} \mathcal{A}_{\rm eff}
\ee
where $n_{\qq_i}$ are the phonon occupation numbers in the initial Fock state $|i\rangle$,
and by considering in eq.~(\ref{eq:Aif}) the six possible intermediate states $|\lambda\rangle$ 
where a virtual phonon is created and reabsorbed (or absorbed and recreated in the six corresponding finite temperature diagrams, in such a way that
the temperature dependence disappears) we find 
\begin{multline}
\mathcal{A}_{\rm eff}(\qq_1,\qq_2;\qq_3,\qq_4)=\frac{1}{16}\sqrt{\frac{\hbar^4\omega_1\omega_2\omega_3\omega_4}{m^4c^8}} \\
\times\left({\rm \Sigma_F}+\frac{(\omega_1+\omega_2)^2 A_{1234}+\omega_{\qq_1+\qq_2}^2 B_{1234}}{(\omega_1+\omega_2)^2-\omega_{\qq_1+\qq_2}^2} \right. \\
+\frac{(\omega_1-\omega_3)^2 A_{1324}+\omega_{\qq_1-\qq_3}^2 B_{1324}}{(\omega_1-\omega_3)^2-\omega_{\qq_1-\qq_3}^2} \\
\left. +\frac{(\omega_1-\omega_4)^2 A_{1423}+\omega_{\qq_1-\qq_4}^2 B_{1423}}{(\omega_1-\omega_4)^2-\omega_{\qq_1-\qq_4}^2}\right)
\label{eq:A2donne2effhydro}
\end{multline}
We introduced the angle-dependent coefficients
\bea
A_{ijkl}&=&(3\Lambda_F+u_{ij})(1+u_{kl})\notag\\
&+&(3\Lambda_F+u_{kl})(1+u_{ij})+(1+u_{ij})(1+u_{kl}) \\
B_{ijkl}&=&(3\Lambda_F+u_{ij}) (3\Lambda_F+u_{kl})
\eea
with $u_{ij}=\qq_i \cdot \qq_j/q_i q_j$
and the thermodynamic quantities
\bea
{\rm \Sigma_F} &\equiv& \frac{\rho^3}{mc^2}\frac{\dd^3\mu}{\dd\rho^3} \\
\Lambda_{\rm F}  &\equiv& \frac{\rho}{3}\frac{\dd^2\mu}{\dd\rho^2}\bb{\frac{\dd\mu}{\dd\rho}}^{-1}
\eea
From the coupling amplitude $\mathcal{A}_{\rm eff}$ we finally obtain the sought low-energy effective Hamiltonian for the $2\leftrightarrow 2$
phonon process in a cubic quantization volume of size $L$: 
\be
\hat{H}_{\rm eff} = \frac{mc^2}{\rho L^3} \!\!\!\! \sum_{\substack{\qq_1,\qq_2,\qq_3,\qq_4 \\ \qq_1+\qq_2=\qq_3+\qq_4}}  \!\!\!\!  \mathcal{A}_{\rm eff}(\qq_1,\qq_2;\qq_3,\qq_4)   \hat{b}_{\qq_3}^\dagger \hat{b}_{\qq_4}^\dagger \hat{b}_{\qq_1} \hat{b}_{\qq_2} 
\label{eq:H2donne2eff} 
\ee

\section{The phonon damping rate}

To calculate the phonon damping rate from the effective Hamiltonian \eqref{eq:H2donne2eff} 
we use as in reference \cite{Sinatra2007} a master equation approach for a harmonic oscillator $\hat{b}_\qq$, $\hat{b}_\qq^\dagger$
coupled to a thermal bath, see complement $\mathrm{B}_{\rm IV}$ in reference \cite{Cohen}:
the phonon mode of momentum $\hbar\qq$ is linearly coupled to the thermal 
reservoir containing all the other phonon modes 
by an interaction Hamiltonian $\hat{H}_{\rm int}=\hat{R}^\dagger  \hat{b}_\qq + \hat{R} \hat{b}_\qq^\dagger$,  where 
$\hat{R}^\dagger$ collects terms in $\hat{b}_{\qq_3}^\dagger \hat{b}_{\qq_4}^\dagger \hat{b}_{\qq_2}$
from $\hat{H}_{\rm eff}$. With respect to Fermi-golden-rule-based rate equations 
or linearised kinetic equations as in reference \cite{Khalatnikov1949b}, the master equation
gives a complete picture of the quantum dynamics of the phonon mode $\qq$, including the case to come of an initial Glauber
coherent state, while of course giving the same damping rate for the phonon occupation number. 
In the Born-Markov approximation \cite{Cohen}, the master equation for the density operator 
$\hat{\sigma}$ of the phonon mode $\qq$ takes the usual form \cite{Cohen}:
\begin{multline}
\frac{\dd}{\dd t}\hat{\sigma}=
\frac{1}{\ii \hbar} [\hbar\omega_\qq \hat{b}_\mathbf{q}^\dagger
\hat{b}_\mathbf{q}, \hat{\sigma}]
+ \Gamma_{\qq}^{-} \hat{b}_\mathbf{q} \hat{\sigma} \hat{b}_\mathbf{q}^\dagger
+ \Gamma_{\qq}^{+} \hat{b}_\mathbf{q}^\dagger \hat{\sigma} \hat{b}_\mathbf{q} \\
-\frac{1}{2}\left\{\Gamma_{\qq}^{-} \hat{b}_{\mathbf{q}}^\dagger \hat{b}_{\mathbf{q}}
+\Gamma_{\qq}^{+} \hat{b}_\mathbf{q} \hat{b}_{\mathbf{q}}^\dagger,\hat{\sigma}\right\}
\end{multline}
where $[,]$ and $\{,\}$ are the commutator and the anticommutator of two operators, and the raising and lowering
rates $\Gamma_\qq^{\pm}$ are given as in \cite{Cohen} (see also equations (A6) and (A7) of \cite{Sinatra2007}) by integrals
of free temporal correlation functions of $\hat{R}$ and $\hat{R}^\dagger$ in the equilibrium state of the reservoir.
As the reservoir is in thermal equilibrium, the two rates are related by $\Gamma_\qq^-=\exp(\hbar\omega_\qq/k_B T)
\Gamma_\qq^+$ \cite{Cohen} and can be expressed in terms of their difference $\Gamma_\qq\equiv \Gamma_\qq^--\Gamma_\qq^+$,
with
\be
\Gamma_\qq=\int_{-\infty}^{+\infty} \frac{\dd t}{\hbar^2} \eee^{-\ii\omega_\qq t} 
\langle [\hat{R}(0),\hat{R}_{\rm free}^\dagger(t)]\rangle_{\rm th}
\label{eq:Gammaqintegral}
\ee
Here $\langle\ldots\rangle_{\rm th}$ is a thermal average, and the free evolution is governed by the quadratic part
of the Hamiltonian, $\hat{R}_{\rm free}^\dagger(t)=\exp(\ii \hat{H}_2t/\hbar) \hat{R}^\dagger(0) \exp(-\ii \hat{H}_2t/\hbar)$.
As explained in \cite{Cohen}, $\Gamma_\qq$ gives both the exponential relaxation rate of the mean occupation number
of the mode $\qq$ towards its thermal equilibrium value $\bar{n}_\qq$
and of the mean phonon amplitude in mode $\qq$ towards zero:
\bea
\frac{\dd}{\dd t} \langle \hat{b}_\qq^\dagger(t)\hat{b}_\qq(t)\rangle &=& -\Gamma_\qq (\langle \hat{b}_\qq^\dagger(t)\hat{b}_\qq(t)\rangle
-\bar{n}_\qq) \\
\frac{\dd}{\dd t} \langle \hat{b}_\qq(t)\rangle &=& -\left(\ii \omega_\qq +\frac{\Gamma_\qq}{2}\right)
\langle \hat{b}_\qq(t)\rangle
\eea
Calculating the commutator in the expression (\ref{eq:Gammaqintegral}) gives
\begin{multline}
\Gamma_\qq=\frac{(mc^2)^2}{4\pi^5\hbar^2\rho^2}
\int  \dd^3q_2 \, \dd^3q_3  |\mathcal{A}_{\rm eff}(\qq,\qq_2;\qq_3,\qq_4)|^2 \\ \times\delta(\omega_3+\omega_4-\omega_2-\omega_{\qq})\bbcro{\bar{n}_2(1+\bar{n}_3+\bar{n}_4)-\bar{n}_3 \bar{n}_4}
\label{eq:Gamma_result}
\end{multline}
where $\qq_4=\qq+\qq_2-\qq_3$ due to momentum conservation and we have used the notation $\bar{n}_i\equiv\bar{n}_{\qq_i}$ to indicate the phonon occupation numbers in thermal equilibrium
\footnote{The expression involving the $\bar{n}_i$ between the square brackets in equation (\ref{eq:Gamma_result}) 
results from the simplification of the extended form $\bar{n}_2(1+\bar{n}_3)(1+\bar{n}_4)-(1+\bar{n}_2) \bar{n}_3\bar{n}_4$.
The extended form is physically more transparent as it reveals the spontaneous and stimulated factors appearing 
in the direct $\qq+\qq_2\to \qq_3+\qq_4$ and reverse $\qq_3+\qq_4\to \qq+\qq_2$ processes. 
Using $(1+\bar{n}_i)/\bar{n}_i=\exp(\hbar\omega_i/k_B T)$ and energy conservation, it can also be rewritten
as in equation (7.4) of reference \cite{Khalatnikov1949b}, that is $(1+\bar{n}_2)\bar{n}_3\bar{n}_4/\bar{n}_\qq$. 
We shall use this last form in equation (\ref{eq:Gammatilde}).}. 
The result (\ref{eq:A2donne2effhydro}),(\ref{eq:Gamma_result}) goes beyond that of Ref.\cite{Khalatnikov1949} where the authors concentrated on a single diagram for the intermediate processes. This will have a significant impact on the behavior of the damping rate as a function of $q$ as we will see.
To go further analytically, we restrict to sufficiently low temperature 
\be
\epsilon\equiv\frac{k_B T}{mc^2} \ll 1
\ee
so that only the (almost) linear region of the phonon dispersion relation is explored 
and the integral \eqref{eq:Gamma_result} is dominated by configurations in which the four involved wavevectors are almost aligned. By introducing the rescaled quantities 
\be
\tilde{q}_i=\frac{\hbar c q_i}{k_{\rm B}T}  \quad \mbox{and}\quad
\tilde{\theta}_i = \frac{\theta_i}{\epsilon |\gamma|^{1/2}}
\label{eq:eps_tildeq}
\ee
with $\theta_i$ the angles between $\qq_i$ and $\qq$ and $\gamma$ the curvature parameter in the spectrum \eqref{eq:fleur}, and by expanding the coupling amplitude \eqref{eq:A2donne2effhydro}
for $\epsilon \to 0$ at fixed $\tilde{q}_i$ and $\tilde{\theta}_i$, we obtain the central result of this paper:

\begin{figure} 
\begin{center}
\includegraphics[width=0.49\textwidth,clip=]{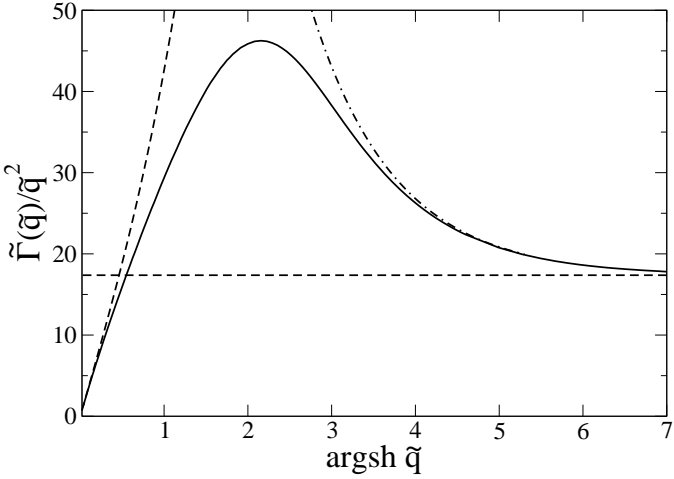}
\end{center}
\caption{ \label{fig:F} Rescaled phonon damping rate due to $2\leftrightarrow2$ phonon processes in a superfluid spin-1/2 Fermi gas, as a function of the rescaled wavenumber $\tilde{q}=\hbar c q/k_BT$, or more precisely of its inverse hyperbolic sine
$\argsh\tilde{q}=\ln(\tilde{q}+\sqrt{1+\tilde{q}^2})$.  
Due to these rescalings, see eq.~(\ref{eq:mainresult}), the result is universal and applies at sufficiently low temperature in the whole region of the BEC-BCS crossover where the dispersion relation is concave at low $q$. The dashed lines show the limiting behaviors  \eqref{eq:petitsq} and \eqref{eq:grandsq} and the dot-dashed curve at large  $\tilde{q}$ is a fit of 
$\tilde{\Gamma}(\tilde{q})/\tilde{q}^2$ by an affine function of $1/\tilde{q}$.}
\end{figure} 

\be
\frac{\hbar \Gamma_{\qq}}{\epsilon_F}  \stackrel{\tilde{q}\,\mathrm{fixed}}{\underset{\epsilon\to0}{\sim}}
\frac{K}{|\gamma|}\bb{\frac{T}{T_F}}^7\tilde{\Gamma}(\tilde{q}) 
\label{eq:mainresult}
\ee
In this formula $\epsilon_F=k_BT_F=\hbar^2k_F^2/2m$ is the Fermi  energy,  $K$ is an interaction-dependent thermodynamic quantity
\be
K = 2 \bb{\frac{3}{4}}^6 \bb{\frac{\epsilon_F}{mc^2}}^3  {(1+\Lambda_F)^4}
\label{eq:Kappa}
\ee
and the rescaled phonon damping rate $\tilde{\Gamma}(\tilde{q})$, shown in fig.~\ref{fig:F}, is a universal, 
monotonically increasing function of the dimensionless wavenumber. Explicitly
\begin{multline}
\tilde{\Gamma}(\tilde{q}) = \int_0^\infty \dd\tilde{q}_2 \int_0^{\tilde{q} + \tilde{q}_2} \dd\tilde{q}_3 \frac{ \tilde{q}_2^3\tilde{q}_3^3(\tilde{q}+\tilde{q}_2-\tilde{q}_3)}{\tilde{q}|v|}   \\
\times \frac{[1+f(\tilde{q}_2)] f(\tilde{q}_3) f(\tilde{q}+\tilde{q}_2-\tilde{q}_3)}{f(\tilde{q})}  \\
 \!\!\!\!\!\! \!\!\!\!\times \int_{0}^\pi \dd\phi \int_{0}^{\pi/2} \dd\alpha\sin\alpha\cos\alpha \, \Theta\bb{-\frac{v}{u}}  \left\vert\mathcal{A}_{\rm red} \right\vert^2 
\label{eq:Gammatilde}
\end{multline}
where $\Theta(x\geq0)=1,\ \Theta(x<0)=0$ is the Heaviside function, $f(x)=1/(\eee^x-1)$ originates from the Bose law, 
$u$ and $v$ are the following functions of $\tilde{q},\tilde{q}_2,\tilde{q}_3,\alpha=\arctan \theta_3/\theta_2$,
and of the relative azimuthal angle $\phi$ of $\qq_2$ and $\qq_3$:
\bea
\!\!\!\!\!\! u &\!\!=\!\!& \frac{\tilde{q}(\tilde{q}_3 \sin^2\alpha - \tilde{q}_2 \cos^2\alpha)+\tilde{q}_2 \tilde{q}_3 (1-\sin2\alpha\cos\phi)}{\tilde{q}+ \tilde{q}_2 - \tilde{q}_3}  \\
\!\! \!\!\!\!v &\!\!=\!\!& \frac{1}{4} \left[ \tilde{q}^3+\tilde{q}_2^3-\tilde{q}_3^3-(\tilde{q}+\tilde{q}_2-\tilde{q}_3)^3 \right]
\eea
We introduced in eq.~\eqref{eq:Gammatilde} the reduced coupling amplitude
\begin{multline}
\mathcal{A}_{\rm r\text{e}d} = 
\frac{1}{\tilde{q}_2\bb{\frac{\cos^2\alpha}{(\tilde{q} +\tilde{q}_2)^2}-\frac{3u}{4v}}}  - \frac{1}{\tilde{q}_3\bb{\frac{\sin^2\alpha}{(\tilde{q} -\tilde{q}_3)^2}-\frac{3u}{4v}}} \\ - \frac{1}{ \frac{\tilde{q}_2 \cos^2\alpha- \tilde{q}_3 \sin^2\alpha+u}{(\tilde{q}_2 -\tilde{q}_3)^2}-\frac{3u}{4v}(\tilde{q}+\tilde{q}_2-\tilde{q}_3)}
\label{eq:Ared}
\end{multline}
The limiting behaviors of the normalized universal Landau-Khalatnikov damping rate \eqref{eq:Gammatilde}  
\bea
\tilde{\Gamma}(\tilde{q}) &\underset{\tilde{q}\to0}{=}& \frac{16\pi^5}{135} \tilde{q}^3 +O(\tilde{q}^4) \label{eq:petitsq}\\
\tilde{\Gamma}(\tilde{q}) &\underset{\tilde{q}\to\infty}{=}& \frac{16\pi\zeta(5)}{3} \tilde{q}^2 + O(\tilde{q}) \label{eq:grandsq}
\eea
are shown in fig.~\ref{fig:F} as dashed lines. Here $\zeta$ is the Riemann zeta function.
Equations \eqref{eq:petitsq}-\eqref{eq:grandsq} disagree with Eqs.(7.6) and (7.12) in reference \cite{Khalatnikov1949}, even for the order in $\tilde{q}$, our results being subleading
by two orders. 
It was already noted in reference \cite{Wyatt1992} that the diagrams neglected by Landau and Khalatnikov
in the calculation of the effective amplitude $\mathcal{A}_{\rm eff}$ are comparable in magnitude with the one that they keep. 
As our explicit calculations show, the failure of this approximation is striking in the low $\tilde{q}$
and in the high $\tilde{q}$ behaviour of $\tilde{\Gamma}(\tilde{q})$, where the neglected diagrams turn out to interfere destructively with
the supposedly leading diagram kept in reference \cite{Khalatnikov1949}, lowering the final leading order in $\tilde{q}$.

There is an interaction strength in the BEC-BCS crossover where the curvature parameter $\gamma$ vanishes, 
since $\gamma$ is positive in the BEC limit \cite{CKS2006} and $\gamma$ is negative in the BCS limit \cite{Strinati1998}. 
Close to this point, on the side $\gamma<0$, one may expect from 
the factor $1/|\gamma|$ in eq.~(\ref{eq:mainresult}) that there is a strong enhancement of the damping rate.
This is however not the case. The result (\ref{eq:mainresult}) can only be used at low enough temperature 
$k_B T < mc^2 |\gamma/\eta|^{1/2}$, such that the cubic term $\hbar c q \frac{\gamma}{8} (\hbar q/mc)^2$
in the dispersion relation (\ref{eq:fleur}) is large as compared to the quintic one $\hbar c q \frac{\eta}{16}(\hbar q/mc)^4$
for typical thermal wave numbers $q\approx k_B T/\hbar c$. For the same reason, one must also have
$\hbar q< mc|\gamma/\eta|^{1/2}$ where $q$ is the wave number of the phonon mode whose damping is under consideration.
As a result, within the validity range of eq.~(\ref{eq:mainresult}), and due to $\tilde{\Gamma}(\tilde{q})=O(\tilde{q}^2)$,
one has $\Gamma_\qq/\epsilon_F = O(|\gamma|^{5/2}/|\eta|^{7/2})$. In practice, at the point
$\gamma=0$ or very close to it, a new calculation should be performed using a linear plus quintic dispersion relation.
Note that, according to the RPA, $\gamma=0$ for $1/k_Fa\simeq -0.144$, in which case $\eta\simeq -0.0428$, 
the dispersion relation remains concave at low $q$ and the Landau-Khalatnikov process is still the leading damping mechanism
\cite{KCS2016}.\footnote{On the side $\gamma$ small and positive, Beliaev-Landau decay is allowed only for wavevectors
$\hbar q< mc[8\gamma/(5|\eta|)]^{1/2}$ \cite{KCS2016}; for higher wave numbers, the quintic term eventually dominates
and the dispersion relation becomes concave. This situation resembles that of superfluid liquid helium-4 \cite{Wyatt2009}.}

\begin{figure} 
\begin{center}
\includegraphics[width=0.49\textwidth,clip=]{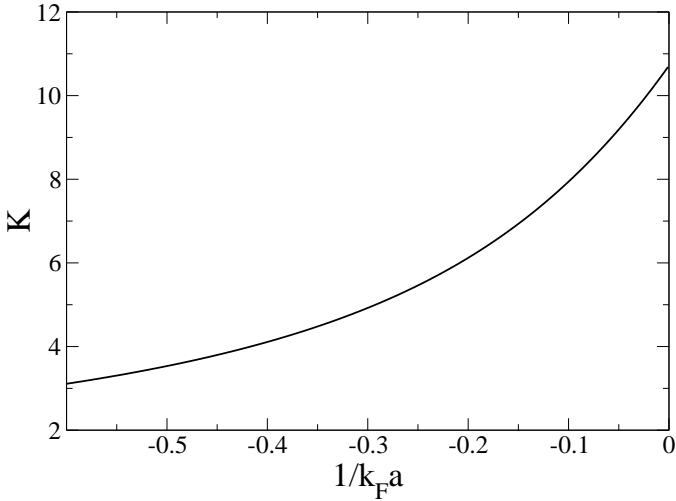}
\end{center}
\caption{ \label{fig:Kappa} Dimensionless thermodynamic quantity \eqref{eq:Kappa} as a function of $1/k_Fa$ where $k_F=(3\pi^2\rho)^{1/3}$ is the Fermi wavenumber and $a$ the $s$-wave scattering length, from the measured zero-temperature equation of state of the spin-1/2 Fermi gas \cite{Salomon2010,TheseNavon}.}
\end{figure} 

\begin{figure} 
\begin{center}
\includegraphics[width=0.49\textwidth,clip=]{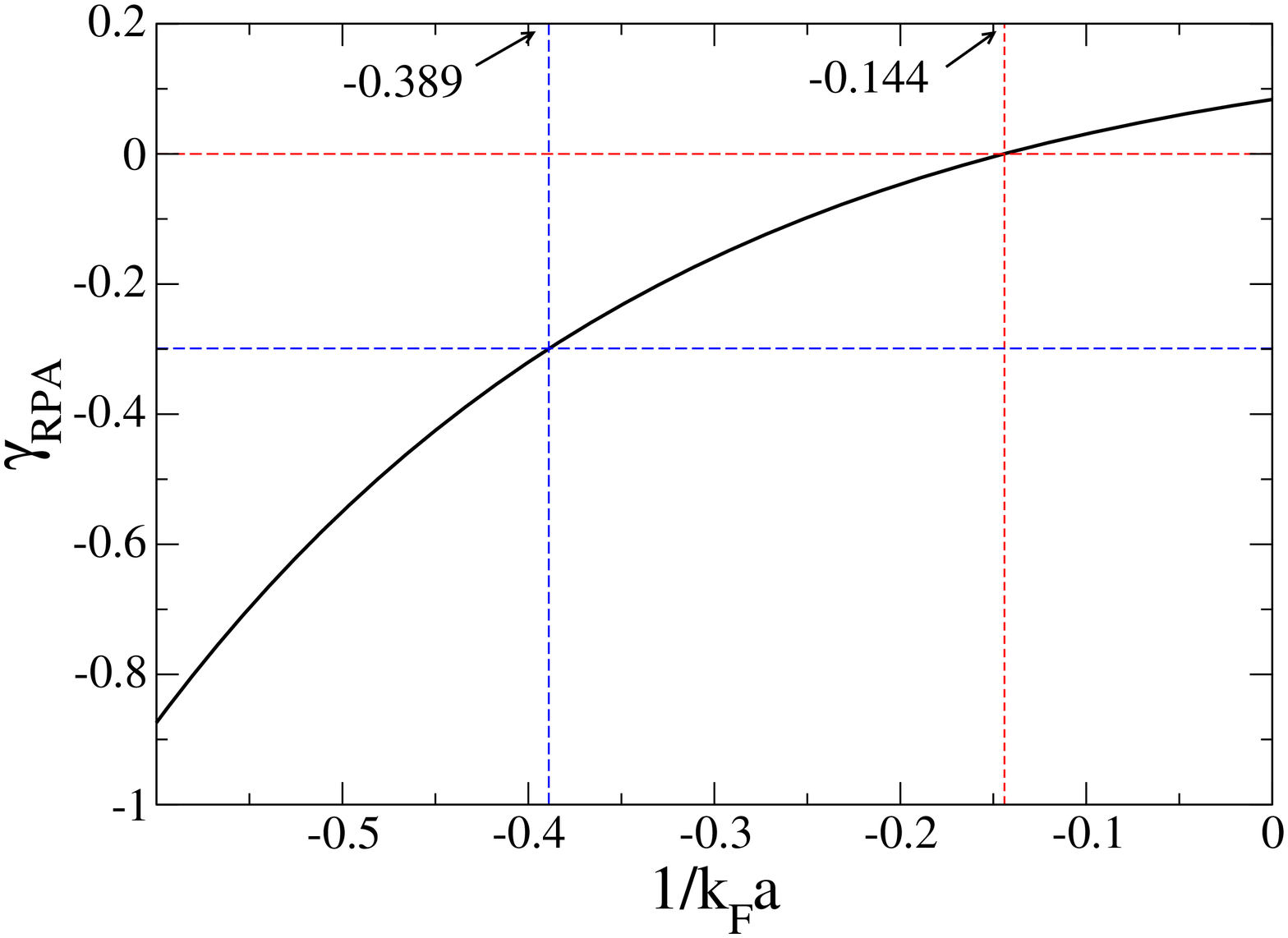}
\end{center}
\caption{ \label{fig:gamma_RPA} Curvature parameter $\gamma$ obtained with the RPA
in reference \cite{KCS2016} for the phonon dispersion relation \eqref{eq:fleur}, as a function of $1/k_Fa$ where $k_F$ is the Fermi wavenumber and $a$ the $s$-wave scattering length.
The dashed lines mark the two points where either the cubic ($1/k_Fa\simeq -0.144$) or the quintic ($1/k_Fa\simeq -0.389$) 
correction to the linear dispersion relation in eq.~\eqref{eq:fleur} vanishes.}
\end{figure} 

\section{Physical discussion and observability}

We now discuss the main result \eqref{eq:mainresult} and the possibility to test it in state-of-the-art experiments with cold fermionic gases. In fig.~\ref{fig:Kappa} we use the measured equation of state of the spin-1/2 Fermi gas \cite{Salomon2010,TheseNavon} to plot the thermodynamic quantity $K$ in eq.~\eqref{eq:Kappa} as a function of $1/k_Fa$.
The curvature parameter $\gamma$ appearing in the phonon dispersion relation \eqref{eq:fleur} has not yet been measured in strongly interacting Fermi gases. We plot in fig.~\ref{fig:gamma_RPA} the prediction of reference \cite{KCS2016} which relies on the Random Phase Approximation and coincides with that of other approximate theories \cite{Strinati1998,Tempere2011}. 
We plot it separately so that, once an experimental value will be available, it can be used in eq.~\eqref{eq:mainresult}.
To give a numerical example, we choose the interaction strength such that the coefficient $\eta$
of the quintic term in the expansion \eqref{eq:fleur} vanishes, $1/k_Fa\simeq -0.39$ according to reference \cite{KCS2016}. From the measured equation of state,
the value of the sound velocity is $c\simeq 0.43 \hbar k_F/m$ and the parameter $\Lambda_\mathrm{F}$ obeys $1+\Lambda_\mathrm{F}\simeq 0.866$, leading
to the thermodynamic constant $K\simeq 4.2$. The predicted value of the curvature parameter is $\gamma\simeq -0.30$ \cite{KCS2016}.
The temperature should be sufficiently low and the wavenumber $q$ much smaller than $2\Delta/\hbar c$ to 
avoid the excitation of the fermionic branch \cite{Ketterle2008,Jin2010}.
By choosing $T=0.073 \,T_F$, larger than the temperature already achieved in reference \cite{Hadzibabic2003}, and $\tilde{q}\equiv \hbar c q/k_B T=5/2$, one obtains 
$\epsilon\equiv k_BT/mc^2\simeq 0.20$, $\hbar q/mc \simeq 0.50$, $q\simeq 0.21 k_F$, $\hbar c q/2\Delta \approx 0.2$
and $\tilde{\Gamma}\simeq 265$. For the typical value $T_F=1\,\mu$K with ${}^6$Li atoms, 
one obtains $2\pi/q \simeq 6\,\mu$m, $\omega_\qq/2\pi\simeq 3.8$ kHz and $c\simeq 2.2$ cm/s.
Eq.~(\ref{eq:mainresult}) then predicts $\Gamma_\qq \simeq 5\,{\rm s}^{-1}$, that is a phonon lifetime $\Gamma_\qq^{-1}\simeq 190$ ms and a mode
quality factor $\omega_\qq/\Gamma_\qq\simeq 4600$.\footnote{As expected, this is in the collisionless regime since the angular frequency $\omega_\qq$ is much larger than the thermalization rate, that we estimate by the damping rate at the typical thermal wavenumber $q_{\rm th}=k_B T/\hbar c$: one finds $\tilde{\Gamma}(1)\simeq26$ and $\Gamma_{\qq_{\rm th}}/\omega_\qq\simeq 2 \times10^{-5}$.
For comparison, we also give 
the Beliaev-Landau damping rate at the unitary limit (where $\gamma_{\rm RPA}\simeq 0.084 >0$), $\Gamma_\qq^{\rm Bel-Lan}\simeq 
140\,\mathrm{s}^{-1}$ for ${}^6$Li atoms with the same values $\epsilon=0.2$, $\tilde{q}=5/2$ and $T_F=1\,\mu$K.}
Quality factors of this order of magnitude have been observed for the transverse monopole mode of an atomic Bose-Einstein condensate
\cite{Dalibard2002}.

The possibility of trapping cold atoms in flat bottom potentials \cite{Hadzibabic2013}
opens the way to a direct test of our prediction in a spatially homogeneous system.
In the box trapping potential, a Glauber coherent state of phonons in a standing-wave mode 
with a well-defined wavevector $\qq$ along a trap axis can be created by laser Bragg excitation of the condensate of pairs 
in the strongly interacting Fermi gas \cite{Vale2008,Hannaford2010,Vale2014}, on top of the preexisting background of thermal phonons.
One simply matches the frequency difference and the wavevector difference of two laser standing waves
to the angular eigenfrequency $\omega_\qq$ and the wavevector $\qq$ of the desired phonon mode.
The subsequent decay of the phonon coherent state
can be monitored by measuring {\sl in situ} the spatial modulation of the density at wavevector $\qq$ using the bosonizing imaging techniques
of reference \cite{Ketterle2005}. Note that in Bragg spectroscopy the density is usually determined after a time-of-flight which amounts to making a measurement
in Fourier space. This is appropriate for $q\gg k_F$ where the scattered atoms separate from the Fermi sea of unscattered atoms.
Here on the contrary $q\ll k_F$ and the measurement is best performed in real space.
In weakly interacting Bose gases, damping times of a fraction of a second have been measured in experiments without being affected by extraneous damping
mechanisms \cite{Dalibard2002}, and the Bragg technique has allowed to measure the Bogoliubov dispersion relation \cite{Davidson2002}
and to observe the zero-temperature $1\to 2$ Beliaev damping \cite{Davidson2002b} with a successful comparison of their $q$-dependence
to theory.

To be complete, let us analyse in more detail the experimental proposal. In reference \cite{Hadzibabic2013} the flat bottom trap has an elongated  cylindric shape.
For simplicity, we model it by an infinite square well potential in the three dimensions with widths $L_x=L_y\equiv L_\perp < L_z$ of rounded up values
$L_\perp=50\,\mu$m and $L_z=100\,\mu$m.\footnote{One may wonder if this is large enough for our infinite-system theory to apply. This will be the case if the typical spacing
between the discrete values of $\omega_3+\omega_4-\omega_2-\omega_\qq$ in the argument of the Dirac distribution in eq.~(\ref{eq:Gamma_result}) 
is $\ll \Gamma_\qq/2$. In the decay
process of an unstable state, the energy is indeed conserved within $\pm \hbar \Gamma_\qq/2$. We estimate the typical spacing by $1/\rho_{\rm states}(\omega_\qq)$
with the density of states 
\begin{equation*}
\rho_{\rm states}(\omega) = \sum^{\rm typ}_{\qq_2,\qq_3} \delta(\omega_3+\omega_4-\omega_2-\omega)
\end{equation*}
where the typical values of $\qq_2$ and $\qq_3$ are at an angle at most $\epsilon$ with respect to $\qq$. Also we impose $q_2< \langle q\rangle_{\rm th}$
where $\hbar c \langle q\rangle_{\rm th} = \frac{\pi^4}{30\zeta(3)} k_B T$ is the mean thermal energy per phonon. The wavenumber $q_3$ is automatically
limited  by $q+q_2$ as in eq.~(\ref{eq:Gammatilde}). Replacing $\sum_\kk$ by $V\int \dd^3k/(2\pi)^3$ in the thermodynamical limit, and taking the 
$\epsilon\to 0$ limit as in $\Gamma_\qq$, we obtain $\rho_{\rm states}(\omega_\qq) \Gamma_\qq/2 \sim (\bar{L}/L_0)^6$ with $\bar{L}^3=V$ the trap volume
and
\begin{equation*}
L_0=\frac{4\pi \hbar}{3mc} \frac{\hbar k_F}{mc} \epsilon^{-7/3} \pi^{-1/3} (1+\Lambda_{\rm F})^{-2/3} (\tilde{\Gamma} I)^{-1/6}
\end{equation*}
where the integral
\begin{equation*}
I=\int_0^{\frac{\pi^4}{30\zeta(3)}} \tilde{q}_2^2 \dd\tilde{q}_2 \int_0^{\tilde{q}+\tilde{q}_2} \tilde{q}_3^2 \dd\tilde{q}_3
\end{equation*}
\begin{equation*}
\times \int_0^1 R^3 \dd R \int_0^{\pi/2} \sin\alpha\cos\alpha\,\dd\alpha \int_0^{\pi} \frac{\dd\phi}{\pi} \delta(uR^2+v)
\end{equation*}
is evaluated numerically, $I\simeq 2.445$. The thermodynamic limit is reached for $\bar{L}>L_0$. In our numerical example $\bar{L}\simeq 63\,\mu$m is indeed
larger than $L_0\simeq 50\,\mu$m. }  
The phonon mode functions are then products of sine functions 
$\frac{2^{3/2}}{V^{1/2}}\prod_{\alpha=x,y,z} \sin(q_\alpha r_\alpha)$, where $q_\alpha L_\alpha/\pi=n_\alpha\in\mathbb{N}^*$ and $V=L_\perp^2 L_z$ is the trap volume. 
During the short time interval $0<t<\tau$, two retro-reflected far-off-resonant Bragg laser beams illuminate the trapped gas.
They induce a conservative lightshift potential $W(\rr,t)=W_0 |\mathcal{E}(\rr,t)|^2$ where $\mathcal{E}(\rr,t)$ is the reduced, dimensionless
positive-frequency part of the laser electric field. In the second quantized form this gives rise to the phonon-light coupling
Hamiltonian 
\be
\hat{H}_W=\int \dd^3r W(\rr,t) \delta\hat{\rho}(\rr)
\ee
Following the values of the physical parameters given above, we take the phonon mode to be excited in the transverse ground state $n_x=n_y=1$ with a wavenumber 
$q_z\simeq 0.5 m c/\hbar$ along $z$, that is $n_z=33$. To optimally excite this mode,
we choose $\mathcal{E}(\rr,t)=\sin(\kk_1\cdot\rr) \eee^{-\ii \omega_1 t}+\ii \cos(\kk_2\cdot\rr) \eee^{-\ii\omega_2 t}$. Here $\kk_i$ and $\omega_i$, the wavevectors and
the angular frequencies of the laser standing waves, obey $\kk_2=\kk_1+q_z\eeee_z$ and $\omega_2=\omega_1-\omega_\qq$, with $\omega_\qq$ 
the angular frequency of the phonon mode and $\eeee_z$ the unit vector along $z$.
In practice the wavevectors $\kk_1$ and $\kk_2$ are in the optical domain, with submicronic wavelength, so $k_1\simeq k_2\gg q$ and the condition $\kk_2=\kk_1+q_z\eeee_z$ 
is achieved by introducing a small angle between them.
In the resulting optical potential 
\begin{multline}
W(\rr,t)=W_0\left\{1-\frac{1}{2}[\cos(2\kk_1\cdot\rr)-\cos(2\kk_2\cdot\rr)] \right. \\
\left. +\phantom{\frac{1}{1}}[\sin(q_z z) -\sin((\kk_1+\kk_2)\cdot\rr)]\sin(\omega_\qq t)\right\}
\end{multline}
the time-independent part is non-resonant and can be neglected. In the time dependent part, the second sine term excites a mode of wavevector $\kk_1+\kk_2$ far from
resonance and can also be omitted. One is left for $0<t<\tau$ with the effective Bragg Hamiltonian \footnote{As the lightshift potential $W_0 \sin(q_z z)\sin(\omega_\qq t)$
is $x$- and $y$-independent, it actually couples to several transversally excited phonon modes during the Bragg pulse. Only the even $x$-parity and even $y$-parity states
are populated, $n_x=2s_x+1$ and $n_y=2 s_y+1$, $(s_x,s_y)\in \mathbb{N}^2$, with normalised amplitudes $c_{s_x,s_y}=2L_\perp^{-2}\int_0^{L_\perp} 
\dd x \dd y \sin \frac{x\pi n_x}{L_\perp} \sin \frac{y\pi n_y}{L_\perp}=\frac{8\pi^{-2}}{n_x n_y}$. 
In the main text we have presented a monomode calculation assuming $|c_{0,0}|^2=1$ as if only the transversally fundamental mode was excited.
In reality, the result in eq.~(\ref{eq:resnaif}) has to be weighted by $|c_{0,0}|^2$ and the transversally excited modes give additional
time dependent contributions to the integrated density modulation that dephase in a time of the order of $t_\perp=q L_\perp^2/(2\pi^2c)\simeq 6$ ms, that is $\simeq 140/\omega_\qq$ and $\simeq 0.03/\Gamma_\qq$.
Since $\sum_{s_x,s_y} |c_{s_x,s_y}|^2$ converges rather rapidly to unity, the relevant $n_x$ and $n_y$ are indeed $\ll n_z$, and the angular frequency
differences $\omega_{(n_x,n_y,n_z)}-\omega_{(1,1,n_z)} \simeq t_\perp^{-1} [s_x(s_x+1)+s_y(s_y+1)]$, with the obvious notation $\omega_\qq=\omega_{(1,1,n_z)}$, etc.
To summarize, eq.~(\ref{eq:resnaif}) becomes $\bar{\delta \rho}(z,t)=\frac{\pi W_0 \rho L_\perp^2}{2mc^2}\sin(q_z z) \{|c_{0,0}|^2 \cos(\omega_\qq t) 
+\re[f(t)\exp(-\ii\omega_\qq t)]\}\eee^{-\Gamma_\qq t/2}$ 
with the periodic function 
\begin{equation*}
f(t)=\sum_{(s_x,s_y)\in\mathbb{N}^{2*}} |c_{s_x,s_y}|^2 \eee^{-\ii [s_x(s_x+1)+s_y(s_y+1)] t/t_\perp}
\end{equation*}
Since $\omega_\qq t_\perp\gg 1$, $f(t)$ plays the role of an envelope function and mainly its modulus matters. 
For our square well potential $|c_{0,0}|^2\simeq 0.657$ and  $|f(t)|<0.343$. This shows that the contributions of the excited modes and their Hamiltonian dephasing are rather harmless and cannot imitate a complete damping of the phonon mode.
}
\be
\hat{H}_W^{\rm res}\simeq 2\hbar\Omega (\hat{b}_\qq+ \hat{b}_\qq^\dagger)\sin(\omega_\qq t)
\ee
of Rabi frequency $\Omega=W_0[\omega_\qq\rho V/(16 \hbar mc^2)]^{1/2}$. Integrating the Heisenberg equations of motion for the phonon annihilation operator
$\hat{b}_\qq$ and taking the expectation
value in the initial $t=0$ thermal state\footnote{A similar calculation, but with bosons and periodic boundary conditions, was done in reference \cite{Cartago}.}, 
one finds $\langle \hat{b}_\qq(\tau)\rangle=\Omega [\tau\eee^{-\ii\omega_\qq\tau}-\sin(\omega_\qq\tau)/\omega_\qq]$ 
where the second term, which is the off-resonant effect of the negative-frequency part of $\sin(\omega_\qq t)$, is zeroed by the choice $\tau=\pi/\omega_\qq\simeq 130 \mu$s.
At the end of the Bragg pulse, this gives a $z$-modulation of the gas density that will subsequently decay to zero due to scattering of thermal phonons.
Since absorption imaging gives in general access to an integrated density, we give the density modulation integrated along $x$ and $y$:
\be
\bar{\delta \rho}(z,t) \underset{t>\tau}{=}\frac{\pi W_0 \rho L_\perp^2}{2mc^2}\sin(q_z z) \cos(\omega_\qq t) \eee^{-\Gamma_\qq t/2}
\label{eq:resnaif}
\ee

\section{Conclusion}

We have calculated the  Landau-Khalatnikov low-temperature phonon damping rate in strongly interacting superfluid Fermi gases. Our expression, thanks to an appropriate rescaling of the wavenumber, takes a universal functional form, plotted in fig.~\ref{fig:F}, that applies in the whole region of the BEC-BCS crossover where the phononic dispersion relation has a concave start. This includes the strongly interacting regime and paves the way to an experimental observation with ultracold atomic gases. 
An alternative, not discussed in detail here, is to use superfluid liquid helium-4 at high pressure, whose
excitation branch is concave at low $q$, while it is convex at low pressure \cite{Wyatt1975}.
The observation of the Landau-Khalatnikov phonon damping would be unprecedented and  would open a new era in the exploration of low-temperature dynamics of macroscopically coherent quantum many-body systems.


\providecommand*\hyphen{-}

\end{document}